\def\pscm{\hbox{$\cm^{-2}\,$}}
\def\pccmK{\hbox{$\cm^{-3}\K$}}
\def\cm{{\rm\thinspace cm}}

\def\erg{{\rm\thinspace erg}}

\def\K{{\rm\thinspace K}}

\def\km{{\rm\thinspace km}}

\def\Msun{\hbox{$\rm\thinspace M_{\odot}$}}

\def\s{{\rm\thinspace s}}

\def\sr{{\rm\thinspace sr}}

\def\ergpcmps{\hbox{$\erg\cm^{-3}\s^{-1}\,$}}
\def\ergpcmsqps{\hbox{$\erg\cm^{-2}\s^{-1}\,$}}

\def\kmps{\hbox{$\km\s^{-1}\,$}}

\def\pcmK{\hbox{$\cm^{-3}\K$}}

\def\psqcm{\hbox{$\cm^{-2}\,$}}

\documentstyle[psfig]{mn}
\begin{document}
\hsize=6truein

\title{The physical conditions within dense cold clouds in cooling flows II}

\author[]
{\parbox[]{6.in} {G.J. Ferland$^{1,2}$, A.C.~Fabian$^3$ and  R.M. Johnstone$^3$ \\
\footnotesize
1. Department of Physics, University of Kentucky, Lexington, KY 40506, USA \\
2. Canadian Institute for Theoretical Astrophysics, University of Toronto,
McLennan Labs, 60 St George Street, Toronto, M5S 3H8, Canada \\
3. Institute of Astronomy, Madingley Road, Cambridge CB3 0HA \\ }} 

\maketitle

\begin{abstract} This is a progress report on our numerical simulations of conditions in the
cold cores of cooling flow condensations.  The physical conditions in
any non-equilibrium plasma are the result of a host of microphysical
processes, many involving reactions that are research areas in
themselves.  We review the dominant physical processes in our
previously published simulations, to clarify those issues that have
caused confusion in the literature.  We show that conditions in the
core of an X-ray illuminated cloud are very different from those found
in molecular clouds, largely because carbon remains substantially
atomic and provides powerful cooling through its far infrared lines.
We show how the results of the Opacity Project have had a major impact
on our predictions, largely because photoionization cross sections of
atoms and first ions are now calculated to be far larger than
previously estimated. Finally we show that the predicted conditions
are strongly affected by such complexities as microturbulence or the
presence of small amounts of dust. Large masses of cold dense gas, in
addition to the warmer molecular gas detected recently, could be
present in cooling flows.

\end{abstract}

\begin{keywords} Molecular Processses -- atomic processes -- galaxies:
clustering -- cooling flows -- intergalactic medium

\end{keywords}

\section{INTRODUCTION} 

Ferland, Fabian, and Johnstone (1994; hereafter FFJ) computed the thermal
and ionization structure of a constant pressure cloud embedded within a
cooling flow and exposed to radiation from the cooling flow and the cosmic
background. By hypothesis the cloud was taken to be free of dust grains. FFJ
also computed a conventional dusty Orion PDR model as an appendix.  FFJ
tried to identify the most important physical processes affecting the state
of the gas.  Any cloud will cool down to the temperature of the cosmic
background once it is sufficiently shielded from other sources of radiation.
In their calculation they found this occurred after a hydrogen column
density of $4\times 10^{21}\pscm$.

Several papers have questioned this result, mainly working by analogy
with galactic molecular clouds (Voit \& Donahue 1995, hereafter VD;
O'Dea et al 1994; Braine et al 1995; Henkel \& Wiklind 1997). Here we
reconsider the FFJ calculation and go over some details that may have
caused confusion. Any numerical simulation of a non-equilibrium plasma
is the result of a balance between a host of microphysical processes.
Many of these processes are research areas by themselves, and can have
substantial uncertainties. In some circumstances the final results may
be sensitive to an especially effective creation or destruction
mechanism, and in others to dozens or more channels. On top of this
there may be fundamental questions such as whether the gas is in
steady state.

The purpose of the present paper is to investigate in detail what
happens to a cloud which is maintained at a reasonably high pressure
($\sim 10^5\pcmK$) for billions of yr and at the same time is
continuously exposed to X-radiation from the surrounding,
pressure-confining medium. In particular we investigate the physical
processes for the coldest parts of the cloud, which enable the
temperature to be significantly less than 17~K. Small amounts of
microturbulence, or grains, cause it to be much colder still.
Calculations by Puy, Grenacher \& Jetzer (1999) have also found a low final
temperature for a fully molecular cloud in a cooling flow.

\section{What happened in the FFJ model}

Several questions have arisen concerning the calculation presented by
FFJ: a) why does the cooling function not agree with existing
molecular cloud cooling functions, b) does the cooling due to the [CI]
370 and 610 $\mu$m lines violate fundamental thermodynamic limits, and
c) VD point out that the heating efficiencies, taken from Shull and
van Steenberg (1985) have been superceded by more recent calculations
(Xu \& McCray 1991), which do not agree for very low electron
fractions.

In this section we use an early version of CLOUDY (C84.09) to
recompute the model shown in FFJ to illustrate some of the dominant
physics.  We believe that this is the version used for that work.

\subsection{Cooling flow cloud vs molecular cloud cooling functions}

The cooling functions computed for molecular clouds by Goldsmith \&
Langer (1978), used by O'Dea et al (1994) and Braine et al (1995), do
not agree with FFJ. In particular, O'Dea et al (1994) predict a
temperature of between 30 and 50~K. Examination of the original papers
reveals the problem. The FFJ model has no dust which greatly reduces
the H$_2$ formation rate. Also carbon remained mostly atomic across
the cloud (Fig 2, FFJ). This was due to rapid charge transfer, CO +
He$^+\rightarrow$ O + C$^+$ + He, well known to be the dominant CO
destruction process in most environments (Tielens and Hollenbach
1985a). He remains partially ionized due to the low electron
density, its large photoelectric cross section at X-ray energies, and
significant abundance of suprathermal electrons from X-ray
photoionization.

This situation is totally unlike that assumed in calculations of the
molecular cloud cooling function.  There, cooling due to atomic carbon is
assumed to be a minor component. For an environment where C/O $<$ 1 little
atomic carbon will exist when the gas becomes totally molecular and all C is
locked up in CO.  The difference between FFJ and molecular cloud cooling
functions is the contribution of [CI] cooling, which can be intense.

VD present separate curves for the cooling due to CO and [CI]; our result
agrees with that of VD for neutral atomic gas, i.e. where the major low
temperature coolant is [CI].

\subsection{The black body limit to [CI] emission}

The [CI] 370, 610$\mu$m lines were especially important coolants in
FFJ's calculation.  Their Fig. 3 shows that these lines dominated the
cooling beginning from a depth of roughly $1.5\times10^{16} \cm$
($3\times 10^{19}\pscm$) and extend to the point where CO was the
dominant coolant.  The temperatures over this region were
below roughly 15~K. The issue here is whether the emergent intensity
of the [CI] lines violates fundamental thermodynamic limits.

For a cloud with constant source function $S_{\nu}$, the emergent
specific intensity is trivially given by $I_{\nu} = S_{\nu} [1-
\exp(-\tau)].$ For a thermalized line, for frequencies near the line
center $S_{\nu}$
is the Planck function at the excitation temperature, and we obtain
the familiar result that the intensity saturates at the black body
intensity. The astrophysical flux ${\cal F}_{\nu}$ (Allen 1973; p90)
is then $\pi B_{\nu}$ and the one-sided emittance, or energy emitted
by a cloud into $2\pi \sr$, will be $\pi B_{\nu} \delta \nu$.

FFJ gave the total emittance or total energy emitted by both sides of a cloud
into
$4\pi \sr$.  The black body limit, for the plane parallel approximation, is
$2\pi B_{\nu} \delta \nu$.  For $\delta \nu$ we take the full
width at half max for the [CI] 610 $\mu$m line.  It has an optical depth of
roughly 80, so the line is optically thick to 2.09 times the Doppler core
(Elitzur \& Ferland 1986).

The cloud computed by FFJ was not isothermal, and it is not possible to
determine at what temperatures the emergent [CI] 610 $\mu$m line was produced,
given the figures shown.  FFJ did show the fraction of the total cooling
carried by the lines, and the lines dominated the cooling at depths between
1.5 and $5\times10^{16} \cm$. It is important to remember that heating and
cooling balance one another, and that the heating rate falls drastically
across the cloud as the incident continuum is attenuated.  The statement
that the line is a uniformly major contributor to the cooling between these
depths is not equivalent to the statement that these depths are the dominant
contributors to the total emittance.

Figs. 1a and 1b show the temperature structure (upper panel) and local emissivity
over the regions where the [CI] fine structure lines form.
The [CI] 610 $\mu$m line
forms predominantly within a narrow region near $1.4\times10^{16} \cm$.  The
emissivity-weighted mean temperature where the 610$\mu$m line is formed is 27.5~K,
and the total emittance
for a black body at this temperature and line width is $1.3\times10^{-5}
\ergpcmps$.  The total emittance predicted by FFJ was substantially less at
$2.1\times 10^{-6} \ergpcmps$.

\begin{figure}
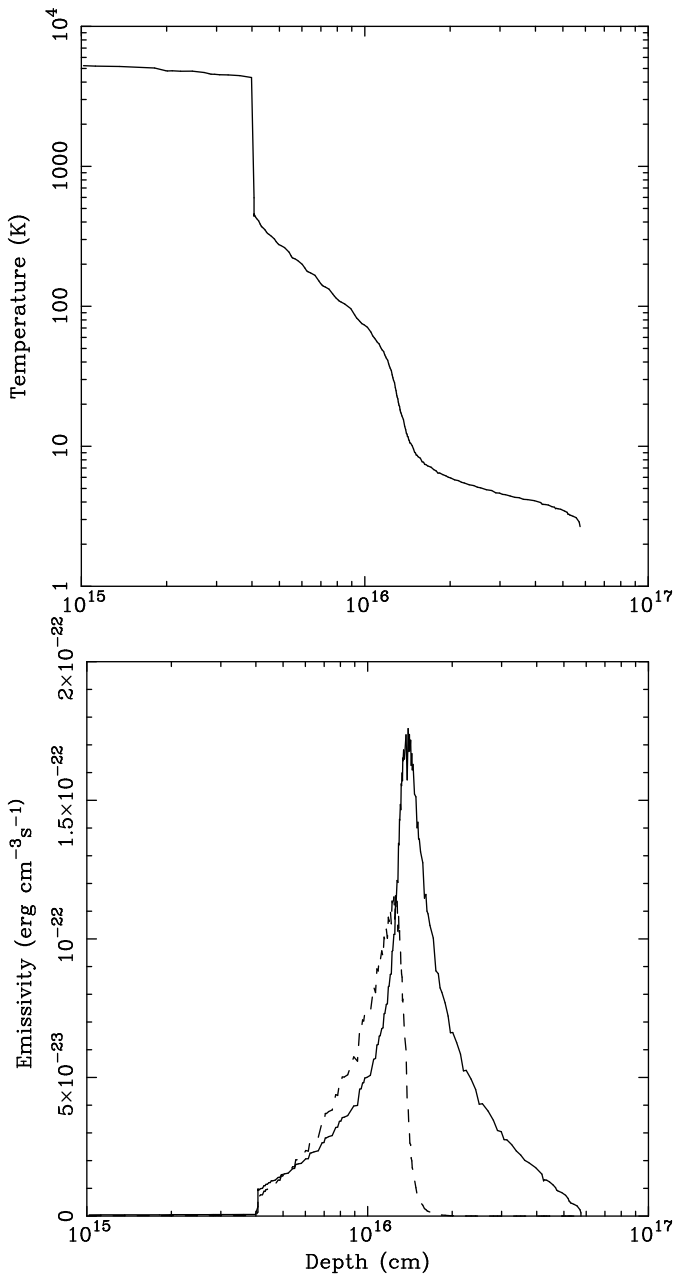

\centerline{\psfig{figure=fig1a.ps,width=0.5\textwidth,angle=270}}
\centerline{\psfig{figure=fig1b.ps,width=0.5\textwidth,angle=270}}
\caption{For the model computed in FFJ, using C84.09 --
a) Upper panel: Electron temperature as a function of depth into the cloud.
b) Lower panel: Emissivity of [CI] 370$\mu$m (dashed), and 610$\mu$m (solid) lines
as a function of depth into the cloud.}
\end{figure}

Although Fig. 1b shows that the emissivity of the 610 $\mu$m line has a tail
that extends to substantially cooler temperatures we confirm that the
emissivity in this line does not exceed the black body limit at any point
within the cloud. 

We note that the geometry assumed by the CLOUDY code is appropriate
for a cloud in the outer parts of a cooling flow (say at 100~kpc
radius, which is that assumed by FFJ). The radiation from the cooling
flow is incident on one face; the other, cold face, is exposed to deep
space and the microwave background. This will be inappropriate for
clouds near the centre of the flow. Since they are heated from all
sides it is plausible that the inner cores of such clouds are warmer
than calculated here and we consider this geometry below. We note that
both cold and warm clouds are commonly seen in the central regions of
cooling flows.

\subsection{Revised low-ionization fraction heating efficiencies}

In an ionized gas all photoelectrons have their energy converted to
thermal energy by rapid elastic collisions with free electrons. In a
predominantly neutral gas, X-ray photoelectrons heat, excite, and
ionize the gas before they are thermalized. FFJ used Shull
\& van Steenberg's (1985)fits to Monte Carlo calculations to take
account of these effects. VD point out that the more recent
calculations of Xu \& McCray (1991) show that these fitting formulae
do not have the proper asymptotic limit. This affects regions with
electron to hydrogen atom ratios less than $10^{-4}.$ Regions of the
cloud modelled by FFJ deeper than $\sim 1.5\times10^{16}\cm$ (or with
a hydrogen column density greater than $3\times 10^{19}\pscm$) do have
electron fractions smaller than this.

Fig. 2 shows that the effects of changing from the Shull and van Steenberg
results to those of Xu and McCray do indeed affect the results by
a modest amount, and make the core of the cloud warmer.

\begin{figure}
\centerline{\psfig{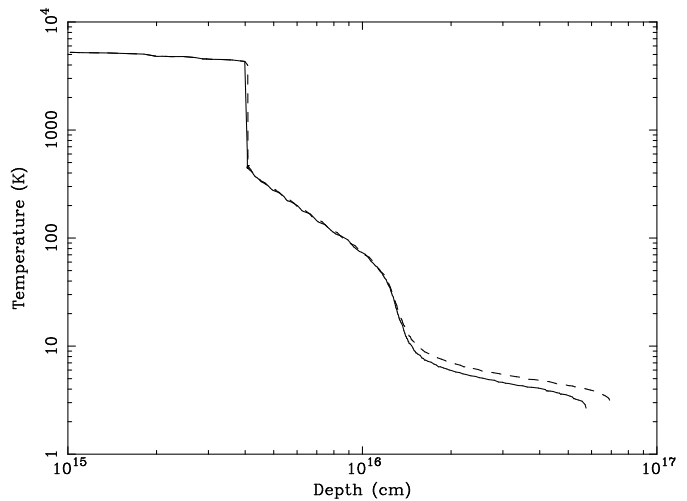}}
\caption{Temperature structure of the constant pressure cloud, using
Shull \& van Steenberg (solid line) and Xu \& McCray (dashed line) X-ray
heating rates, both in C84.09}         
\end{figure}

\section{Advances in the microphysics}

Any simulation of a non-equilibrium plasma rests on a foundation of a
host of microphysical processes.  Estimates of rates and cross
sections have improved enormously over the past decade, some of which
are relevant to the problem addressed here. CLOUDY has been revised to
take account of these advances. Version numbers are used to track the
changes and the CLOUDY home page
http://www.pa.uky.edu/$\sim$gary/cloudy records them in detail.  FFJ
used version C84.09 of the code.  That version had rates and cross
sections that were up-to-date circa 1992.  The current version is C96
and every effort has been made to update the entire atomic/molecular
database to the best current values (Ferland et al 1998).

As we show below, the physical conditions predicted by the two versions of
the code do not agree deep within the cooling flow irradiated cloud.  Here we
identify the physical processes that have caused the computed conditions to
change. We will compare an existing copy of version 84.09 (after correcting
for the proper asymptote for the secondary ionizations) to the version of
the code now available on the web.

\subsection{Constant density model}

The original FFJ calculation was a constant pressure cloud.  This was
motivated by the physical situation, visualized as one in which the hot
X-ray emitting plasma and the cool clouds are in pressure equilibrium.
For this comparison between the two versions we consider constant density
clouds to clarify the real differences in the codes.  The constant pressure
assumption exaggerates differences between similar photoionization models,
since minor differences in the thermal structure are magnified by resulting
changes in the density and ensuing opacity. Otherwise the initial conditions
are identical to FFJ.  In particular we will stop both calculations at the
same column density, $\log(N_H) = 22.8 \psqcm$.  This is the depth at which
the temperature predicted by C84.09 fell below 4K.

\subsubsection{Conditions at the illuminated face of a cloud}

Fig. 3 shows some details of the FFJ standard cloud computed by both
C84.09 and the current version. The electron temperatures at the
illuminated face differ by significant amounts, with the current
version being cooler. Temperature is the result of the balance between
heating and cooling.  Fig. 3 also shows that the photoelectric heating
is nearly the same at the illuminated face.

\begin{figure}
\centerline{\psfig{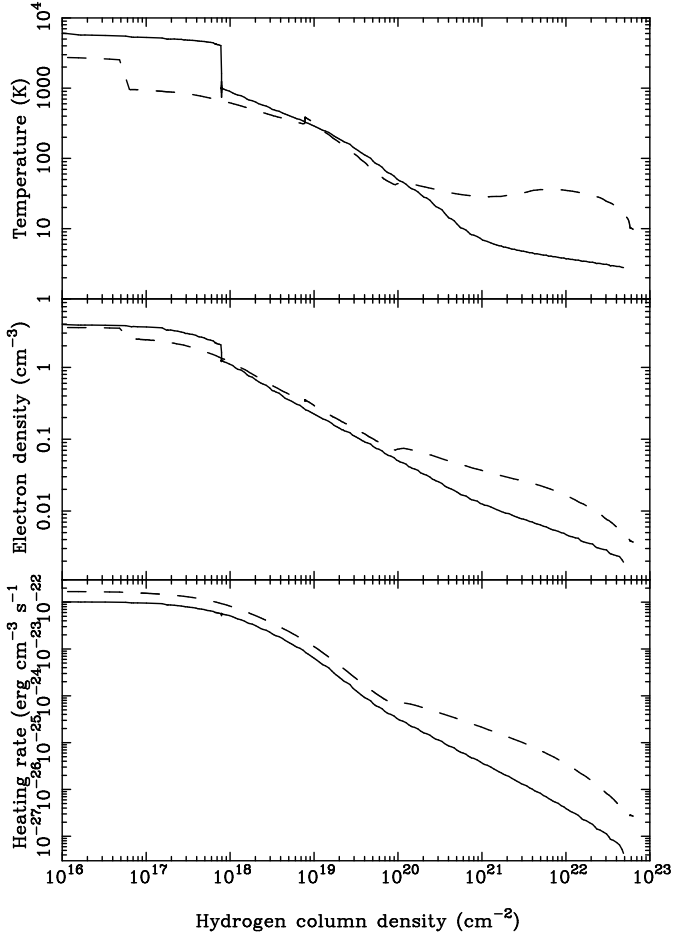}}
\caption{Comparison between cloudy version 84.09 (solid) and 90.04 
Dashed). {\bf NB} These models are for the case of constant density clouds,
(see section 3 for the reason for this) and  not the constant pressure clouds computed by FFJ.}         
\end{figure}

The differences at the illuminated face are caused by changes in the cooling
efficiency of the gas.  The biggest change is due to IR lines within the
ground term of Fe$^+$.  The current version of the code uses collision
strengths from Zhang \& Pradhan (1995).  The [Si II] 34 $\mu$m line is stronger
now; the current version uses the recent collision strengths from Dufton
\& Kingston (1994).  None of these rates were available for C84.09.
Changes in the charge transfer data base, especially reactions
involving He and He$^+$ (Wang J., Stancil, P., Schultz, D., 
Rakovic', M., Kingdon, J.,\&
Dalgarno, A., 2001, in preparation;
http://www-cfadc.phy.ornl.gov/astro/ps/data/), also increased the cooling
by changing the ionization of the gas.  Fe was predicted to be $\sim
50$ per cent in the form of Fe$^+$ in the old calculation, while close
to 100 per cent of Fe is now predicted to be in the form of Fe$^+$.
This further increases [FeII] cooling and results in lower
temperatures. The result of this increased cooling efficiency is that
the gas now equilibrates at a somewhat lower temperature (2700K) than
it did in version 84 (6380K).  The gas pressure is smaller by the
corresponding factor.

The figure also shows that the depth at which the thermal front, where the
gas abruptly changes from the warm ($\sim$4000K) to cool ($<$1000K) phases, is also
different.  This is again due to changes in the details of the cooling
function.

\subsubsection{Conditions deep within a cloud}

Fig. 3 shows that the largest differences occur deep within the cloud.
The photoelectric heating is now roughly a factor of two larger, as is the
electron density and resulting temperature.  These differences are caused by
two major improvements in the atomic database.

Fig. 4 shows the local continuum at the shielded face of the cloud.
The dashed line is the incident continuum produced by the surrounding
cooling flow.  The solid line is the continuum predicted by the later
code, and the dotted line is that predicted by C84.09.  (These
continua are in excellent agreement, and the two are often not
distinguishable.)

\begin{figure}
\centerline{\psfig{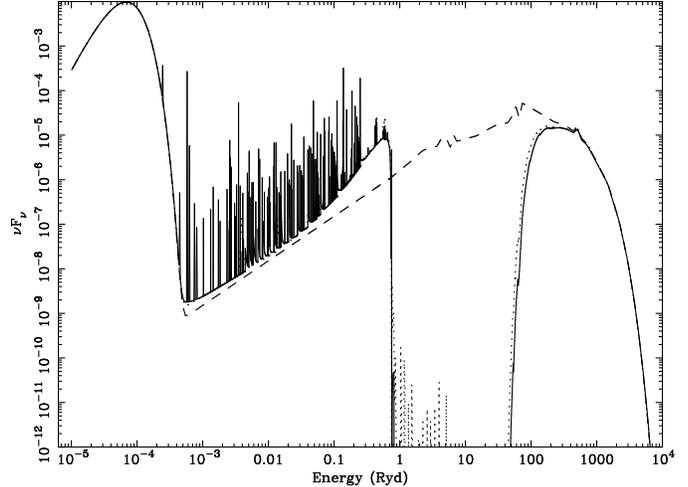}}
\caption{Comparison of spectra for the constant density models. Dashed
line: incident flux. Solid line: flux at shielded face of cloud
predicted by the later code. Dotted line: flux at shielded face of
cloud predicted by C84.09.}
\end{figure}

The continuum is strongly absorbed between the atomic carbon edge (11.2 eV)
and several keV.  Rayleigh scattering provides significant additional
shielding for energies longward of Ly$\alpha$.  The net effect is that the
conditions deep within the cloud are determined by gas interactions with
both very soft (5-10eV) and very hard ($>$ 2keV) radiation.

The entire photoionization cross section database was revised in C90
(Ferland et al 1998). C84.09 used photoionization cross sections
computed by Reilman \& Manson (1979). The physical assumptions they
used were very accurate for highly charged species, but became
increasingly approximate for lower charges.  The database used in C96
is described by Verner et al. (1996).  It uses experimental or Opacity
Project (Seaton 1987) cross sections and explicitly gives partial
cross sections for each subshell.  The Reilman \& Manson and the
Verner et al. data sets are in good agreement for second and more
highly charged ions, and for inner shells of most ions, but are quite
different for valence shells of atoms and first ions. A comparison of
the opacity between 0.1 and 3 Ryd computed by C84.09 and the later
code is shown in Fig. 5.

Deep in the cloud there is actually more power available in reprocessed
Balmer continuum radiation than in the attenuated X-ray continuum (Fig. 4).
This cloud has a very low level of ionization, and there have been dramatic
increases in the photoionization cross sections for neutral atoms and first
ions.  The valence shell photoionization cross sections
for atoms and first ions are now often 0.5 dex to 1.0 dex larger.  Fig. 5
shows that the total gas opacity, as determined by these cross sections and
the resulting ionization balance, also differs by up to 1 dex. This results in
substantially different photoionization rates, photoelectric heating,
and electron densities, mainly due to third row elements with small
ionization potentials (Na, Si, S, Ca). These species were atomic in the
C84.09 solution, but are predicted to be first ions by C96.  This
accounts for roughly half of the differences between the two calculations.

\begin{figure}
\centerline{\psfig{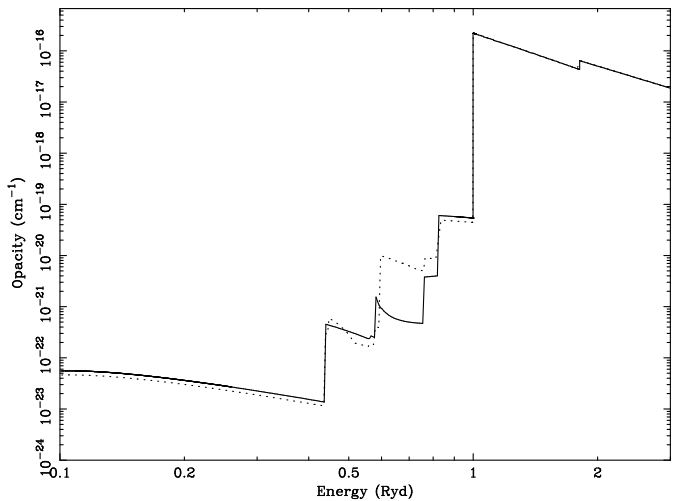}}
\caption{The predicted total opacity at the illuminated face of the cloud for
C84.09 (dashed line) and C96 (solid line).}         
\end{figure}

The remaining differences are largely due to changes in the fluorescent
yields and distribution of Auger electrons.  Version 84 used the approximate
methods outlined by Weisheit \& Dalgarno (1972), Weisheit (1974), and
Weisheit \& Collins (1976).  The current version employs the extensive data
set of Kaastra \& Mewe (1993), combined with detailed subshell
photoionization cross sections (Verner et al. 1996). The distribution of
Auger electrons ejected for third and fourth row elements is significantly
different.  The resulting equilibrium is especially sensitive to details of
inner shell processes since an exceptionally hard continuum is present at
depth (Fig. 4).

The net result is that the gas is more highly ionized with larger
photoelectric heating rates.  It is unusual for a calculation to change by
as much as this one has, because changes in the atomic database tend to be
random and so have a small net effect.  In this case the changes in the
valence shell photoionization cross sections and numbers of Auger electrons
have all been in the sense to result in increased ionization and heating,
and they have had a major net effect.

\subsection{Changes to molecules and grains}

A major effort has gone into making the grain physics state-of-the-art.
The current implementation is described by van Hoof et al.
(2001).  A built-in Mie code can be used to generate optical properties
for any grain constituent, and the grain population can be resolved into
any number of size bins.  Grain charging, heating, temperature and drift
velocity is then computed for each size.  Resolving the grain size
distribution is crucial since smaller grains tend to be hotter, produce
the greatest photoelectric heating of the gas, and so have a profound
effect on the spectrum.  PAHs and single-photon heating (Guhathakurta \&
Draine 1989) are also fully treated.  The current implementation of the
grain physics fully reproduces the results presented by Weingartner \&
Draine (2001) but with the added advantage of including a
self-consistent solution of the physical state of the gas surrounding
the grains (radiation field, electron kinetic energy distribution, etc).

The molecule network presently includes H$^-$, H$_2$, H$_2^+$,
H$_{3}^+$, HeH$^+$, OH, OH$^+$, CH, CH$^+$, O$_2$, O$_2^+$, CO,
CO$^+$, H$_2$O, H$_2$O$^+$, H$_3$O$^+$, and CH$_2^+$.  Reaction rate
coefficients are from Hollenbach and McKee (1979; 1989); Tielens and
Hollenbach (1985a), Lenzuni, Chernoff, and Salpeter (1991), Wolfire,
Tielens, and Hollenbach (1990); Crosas and Weisheit (1993); Puy et al
(1993), Maloney, Hollenbach, \& Tielens (1996), Hollenbach
\& Tielens (1999), and the UMIST database (http://www.rate99.co.uk/).
The resulting chemistry is in good agreement with standard PDR
calculations. The effects of suprathermal electrons are important and
treated as in Dalgarno et al (1999).

CO includes both $^{12}$C$^{16}$O and $^{13}$C$^{16}$O using shielding rates
from van Dishoeck \& Black (1988) and all radiative transfer
processes.  These molecules are treated as rigid rotators with a
complete calculation of the level populations and emission from the
ground rotational ladder. Any number of levels can be included; the
current calculation includes the lowest 50 levels and uses collision
rates from de Jong, Chu, \& Dalgarno (1975). 
The older calculation treated CO rotation cooling using expressions from
Hollenbach \& McKee (1979).  This is crucial for the cloud cores, where
CO is the dominant coolant.  The current calculations obtain the CO
cooling by solving for level populations along the full rotational
ladder, including collisional excitation, deexcitation, continuum
pumping excitation, and line trapping.   This treatment is expected to
be more rigorous because it is evaluated at each point for the detailed
local conditions and line optical depths for lines along the CO
rotational ladder. Tests show that the cooling predicted by the detailed
molecule is generally within a factor of three of that predicted by the
Hollenbach \& McKee approximation.  Besides its greater accuracy, another
benefit of the complete model molecule is that the full rotation
spectrum is predicted. The work of Puy et al 1999 considered the cooling function due to molecules
of H$_2$ CO and HD. We have checked and found that the inclusion of the
HD molecule has no significant effect on our results.

\subsection{Constant pressure model}

Johnstone, Fabian \& Taylor (1998) recalculated the FFJ cold cloud model
using C90.04, and found that there existed an extended region in the core of
the cloud with a temperature of between $13-17$K at column densities up to
$4\times 10^{21}\pscm$ (their Fig. 8). In Fig. 6 we show the temperature
profiles within the clouds for C84.09 (dashed line) and C96 (solid line).
The reasons for the difference in the models between versions are as
explained in section 3.1.

\begin{figure}
\centerline{\psfig{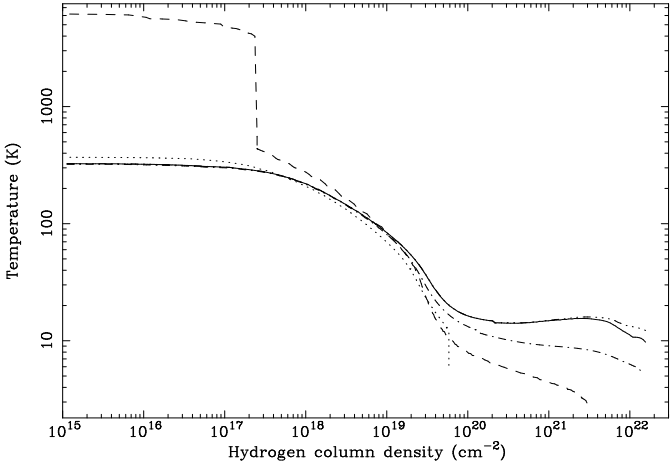}}
\caption{Temperature profiles within the irradiated clouds. All models
assume constant pressure and have the same outer pressure of $2.9\times
10^{5}\pccmK$. Solid line: C96, no microturbulence, no dust. Dashed line:
C84.09, no microturbulence, no dust. Dot-dashed line: C96 with $1\kmps$
microturbulence, no dust. Dotted line: C96, no microturbulence, galactic
dust-to-gas ratio. The dash-dot-dot-dot curve depicts the cloud
illuminated from all sides. }
\end{figure}

\begin{table*}
\caption{Emission line stregths, relative to H$\beta$, predicted
  from the irradiated cloud models.
  Column 1 gives the
  line identifier while column 2 shows the wavelength of the
  line. Columns 3-7 give the line intensities relative to H$\beta$ for
  the cloud irradiated from one side, 
the cloud with 1\kmps microturbulance,
the dusty cloud,
the cloud irradiated from 2 sides, 
and the dusty cloud where the temperature was not allowed to fall
  below 30K,
respectively.
  The surface flux of the H$\beta$ line in units of $\ergpcmsqps$ is
  given for each model at the bottom.
}
\begin{tabular}{lrrrrrr}
&&&&&&\\
Line             &            &1-side &turbulence&dust   &2-side &heat+dust \\
\hline                                                                     
H$\alpha$        &6563 \AA    &5.42   &5.53      &4.21   &5.41   &4.24      \\
P$\alpha$        &1.87 $\mu$m &0.88   &0.87      &0.43   &0.88   &0.84      \\
HeI              &626 \AA     &9.68   &0.56      &1.49   &9.68   &-         \\
{[}CI]           &609 $\mu$m  &11.7   &23.3      &9.33   &11.1   &22.9      \\
{[}CI]           &369 $\mu$m  &28.4   &20.9      &23.2   &25.6   &58.5      \\
{[}CII]          &157 $\mu$m  &3.33   &2.43      &1.3    &3.35   &1.45      \\
{[}OI]           &63.1 $\mu$m &12.8   &12.5      &38.8   &12.8   &39.8      \\
{[}OI]           &145 $\mu$m  &0.43   &0.43      &2.09   &0.43   &2.04      \\
{[}SiII]         &34.8 $\mu$m &15.00  &15.5      &2.48   &15.00  &2.34      \\
{[}FeII]         &25.9 $\mu$m &1.44   &1.48      &0.09   &1.45   &-         \\
$^{12}$C$^{16}$O &1-0         &0.14   &-         &-      &0.13   &0.47      \\
                 &2-1         &0.88   &0.84      &0.02   &0.87   &4.67      \\
                 &3-2         &2.18   &3.5       &0.16   &2.28   &13.1      \\
                 &4-3         &3.02   &5.43      &0.20   &3.42   &15.3      \\
                 &5-4         &2.29   &1.59      &0.11   &3.46   &7.99      \\
                 &6-5         &0.34   &0.02      &0.05   &1.32   &1.92      \\
$^{13}$C$^{16}$O &3-2         &0.1    &0.22      &0.01   &0.14   &0.69      \\
                 &4-3         &0.21   &0.5       &-      &0.38   &0.55      \\
                 &5-4         &0.13   &0.04      &-      &0.35   &0.19      \\
\hline                                                                     
H$\beta$         &Flux        &1.73E-7&1.56E-7   &7.23E-8&1.73E-7&7.62E-8   \\
\hline
\end{tabular}       
\end{table*}

\section{What happens in nature?}

\subsection{Equilibrium timescales}

FFJ's calculations were basically of a conventional photodissociation
region (PDR).  They presented both a grain-free model of a constant
pressure cooling flow cloud and, as an appendix, the dusty Orion blister.
The methods and assumptions they used were based on the chemistry that occurs
in conventional PDR (Tielens \& Hollenbach 1985a,b) and warm shocks
(Hollenbach \& McKee 1979, 1989). Where comparisons were possible the
calculations obtained by FFJ using the then current version of the code
CLOUDY were in reasonable agreement with these papers.
The assumption
that the structure has had time to come into time-steady equilibrium
underlies all of this work.  FFJ did check that the recombination and
thermal timescales were short.

Since that time Draine \& Bertoldi (1996) have shown that, even for
conventional galactic objects like the Orion PDR, the H$_2$ part of
the chemistry is so slow that it may not reach equilibrium (Bertoldi
\& Draine 1996; Draine \& Bertoldi 1996).  The basic reason is that
for a homonuclear species like H$_2$ direct formation is not possible
(there is no dipole moment) and only indirect mechanisms are
available.  In the case of H$_2$ these include radiative association
through H$^-$ and catalysis on grain surfaces.  For a grain-free
environment only the first process is possible.

CLOUDY (C96) now checks timescales for most important parts of the
ionization and thermal solutions. Fig. 7 shows some of these
timescales for the FFJ cloud. This calculation is of a constant
pressure cloud with the density at the illuminated face given in FFJ.
Timescales for recombination, thermal equilibrium, and the formation
of the important species CO and H$_2$, are shown. All timescales are
fast enough for them to reach steady state.  At the shielded face of
the cloud the longest timescale is that for H$_2$ formation, but even
here the timescale of $\sim$1{\rm Gyr} is likely fast enough to have
reached equilibrium. The major effect of the H$_2$ network being out
of equilibrium would be to introduce an uncertainty in the total
hydrogen density since the H and H$_2$ fractions are uncertain. H$_2$
has little influence on the thermal balance of the cloud, however.

\begin{figure}
\centerline{\psfig{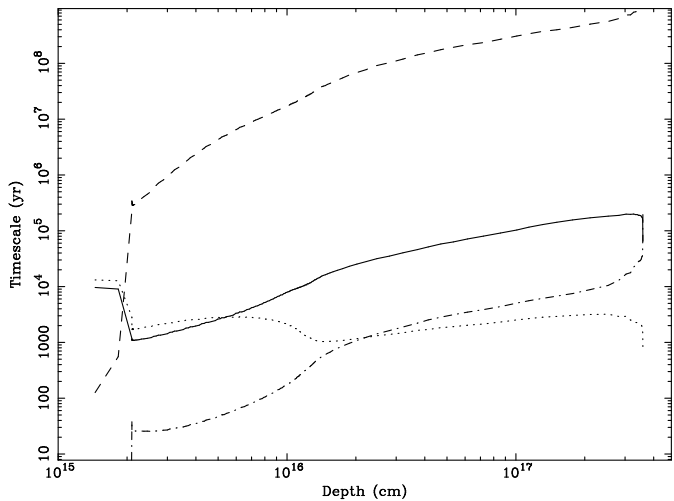}}
\caption{Equilibrium timescales for cooling (solid line), H$_2$ molecule
formation (dashed line), CO molecule formation (dot-dashed line) and
recombination (dotted line) as a function of depth into the cloud.}
\end{figure}

\subsection{Dependence on the velocity field}

One parameter that strongly affects the results is the local velocity
field. The original calculation was of a cloud in pressure equilibrium
with its surroundings.  Only thermal line broadening was assumed.  We
have now computed the temperature structure of the constant pressure
cloud using C96 and including a microturbulent velocity field of $1
\kmps$. (This is the Alfv\'en speed of a matter dominated magnetic
field which has an energy density of only 0.02 per cent of the thermal
energy density at the illuminated face of the cloud. This would be
unusually low in the interstellar medium of our galaxy.) The only
physical effect of the microturbulence is to desaturate the lines and
allow them to cool more efficiently.  The dot-dashed line in Fig. 6
shows the temperature profile resulting from this model; temperatures
are approximately a factor of two cooler at depth. Since the thermal
width for the [CI] fine structure lines is only a fraction of a
kilometer per second, even modest levels of turbulence make a
substantial effect. Most of the core of a realistic, dust-free cloud
is therefore below 10~K.

\subsection{Dependence on the presence of dust}

Grains have a dramatic effect on the structure of these clouds.  Fabian,
Johnstone \& Daines (1994) and more recently Johnstone,
Fabian, \& Taylor (1998) showed the effects of a variety of dust to gas
ratios on the temperature profiles of the clouds. We reproduce the model 
from Johnstone, Fabian \& Taylor which included dust at the galactic
gas-to-dust ratio as the dotted line in Fig. 6.

In the calculations with dust absent the gas had
essentially no opacity in the Balmer continuum.  H$_2$ formation proceeded
only through the H$^-$ route.  At depth the radiation field was dominated by
reprocessed Balmer continuum radiation.  This radiation field dominated the
physical conditions there.

In the dusty model the H$_2$ and CO molecular fractions are far larger than in the grain-free
case; both approach 100 percent  of the H or C abundance.  H$_2$ forms more efficiently
on grain surfaces, and the added
grain opacity shields H$_2$ and CO from photodissociating radiation.  The
grain opacity peaks at energies near 1 Ryd, so the grains get a significant
part of the ionizing radiation field.  The biggest effect of this is that
the reprocessed Balmer continuum is about a factor of 20 weaker in the
dusty case. There is therefore far less heating of the gas at depth and the
cloud reaches the background temperature at a column density of just over
$10^{20}\pscm$.
The [SiII] and [FeII] emission lines are predicted to be much
  weaker than in the non-dusty case due to depletion of these elements
  on to dust grains.

We have considered the effect that the ambient galaxy starlight has on
the structure of our cloud by modelling it as a stellar atmosphere
(Kurucz 1991) with $T_{eff}$=4500K. We set the stellar flux incident on the
cloud by assuming the starlight from the whole galaxy was a point
source with an absolute bolometric magnitude of -23, located at a
distance of 100kpc from the cloud. There is no discernable difference
in the temperature structure of the cloud when this radiation field is
included. We note that on smaller scales, in the central regions of
cooling flow galaxies, there is evidence for excess blue light which
may power the optical emission line regions (Johnstone etal 1987,
Allen 1995, Crawford etal 1995).


\subsection{Cloud irradiated from all sides}

As mentioned above, the models we consider are appropriate for
condensations in the outer regions of the flow, where diffuse emission
from the hot gas strikes only one face of the cloud we model.  The
shielded face is exposed only to the extragalactic background radiation.
The cloud is able to freely radiate in this outer direction, which is
also the coldest part of the cloud.  

For regions closer to the center of the flow, a cloud will be
illuminated from all sides.  In this case the most shielded region
will be the cloud's core, which will not be able to radiate very
efficiently due to the shielding effects of the surrounding gas.  How
does this warmer layer of gas affect the core temperature?  As VD
point out, in the simple case of a cloud cooled only by one optically
thick line, the temperature of the core could not fall below that of
surrounding warmer layer.  Although this is not the situation in the
clouds we consider (there are many different coolants, and different
coolants operate at different depths; FFJ assumed one-sided
illumination), the geometry does occur in parts of the flow.

We did tests to simulate the two-sided illuminated case.  In this case
the cloud core "sees" both the part of the layer we model, and also a
mirror image that is symmetric about the core.  The main effect is that
the total line optical depths are twice as large as those to the
midplane - at the core the line optical depth is the same in all
directions and equal to the optical depth from the illuminated face to
the shielded face in the previous calculations.  

The predicted structure is shown as the dot-dot-dot-dashed line in Fig. 6.
As expected, the clouds are indeed slightly warmer.  The main coolant in
the core is CO, and Table 1 shows that, although the intensity of
the lowest 1-0 transition hardly changes, the higher rotation lines
become somewhat brighter.  As the lower J lines become more optically
thick their upper level population increases, and higher transitions
in the rotation ladder carry the cooling.  These lines, which are the
most efficient coolants, have small optical depths.

\subsection{A heated cloud with dust}

Galactic molecular clouds are not heated by starlight photoionization.
Rather a variety of agents, many involing mechanical or magnetic
processes, act to sustain the temperature at higher-than-expected
levels.  Were this to occur in the clouds we computed, the CO would be
stronger than we predict.  

As a test we recomputed our dusty (zero turbulence) model, but not
allowing the temperature to fall below 30 K.  The full thermal solution
was performed when the equilibrium temperature was found to be above
this limit, but a constant 30 K temperature was used for gas that
otherwise would have been cooler.  The results are given in the last
column of Table 1.  

The hydrogen lines are fainter due to absorption of the incident
continuum by grains, and the Si and Fe lines are fainter due to
depletion of these elements on to grains. The CI lines, and
especially the CO rotation transitions, are however, stronger. 
This is due to the warmer temperature in
regions where they are formed.  Actually the CO lines could increase in
intensity, almost without limit, were the cloud column density to be
increased (the calculation was stopped at a somewhat arbitrary depth).
The important results are the line ratios.

\subsection{Observational evidence for molecular gas and dust in cooling flows}

Until recently, there was little evidence for cold molecular gas in
cooling flows. NGC1275 in the centre of the Perseus cluster had been
detected (Bridges \& Irwin 1998) and many non-detections reported
(Grabelsky \& Ulmer 1990; McNamara \& Jaffe 1992; O'Dea et al 1994;
Braine \& Dupraz 1994; Henkel \& Wiklind 1997). Now CO line emission
has been found in the central galaxies of 16 central cluster galaxies
(Edge 2001). The molecular gas implied by these detections has a
temperature consistent with 20--40~K and masses which range from
$10^9\Msun$ in the weakest objects to $10^{11.5}\Msun$ in the
strongest. More cooler gas could be present; the CO flux predicted
by our standard model
from a square kpc of our irradiated clouds is less than one per cent
of that detected by Edge (2001). Smaller masses of much hotter
molecular gas has also been found in cooling flows through vibrational
H$_{2}$ emission (Jaffe \& Bremer 1997; Falcke et al 1998; Donahue et al 2000; Jaffe, Bremer \& van der Werf 2001; Edge
et al 2002; Wilman et al 2002).

Dust too appears to be common in this environment, and is detected
either by its effect on emission-line ratios (Hu 1992; Donahue \& Voit
1993; Allen et al 1995; Allen 1995; Hansen, Jorgensen \&
Norgaard-Nielsen 1995; Crawford et al 1999), dust lanes (Sparks,
Macchetto \& Golombek 1989; McNamara et al 1996; Pinkney et al 1996)
or FIR/sub-mm emission (Lester et al 1995; Cox, Bregman \& Schombert
1995; Allen et al 2001; Edge 2001; Irwin, Stil \& Bridges 2001).

The origin and heating of the molecular gas and grains is currently
unknown. They may form from gas cooled from the cooling intracluster
medium and be enriched and photoionized by the emissions of massive
young stars formed within the cooled clouds, or they may have another
source. This uncertainty remains also for the optical nebulosity common
in cooling flows. The cold clouds modelled in this paper would only be
detectable in emission if they are both dust free and have a
covering fraction exceeding 10 per cent over the telescope beam
(scaling from the
results of Edge 2001). Such a high covering fraction would imply a
molecular mass greater than $10^{11}\Msun$ for objects at $z\sim0.1$
and the instruments used by Edge (2001).
However, part of the emission seen may
be due to the clouds modelled here.
Detection by absorption would, of
course, depend on the covering factor and velocity spread of the
cloud population.

\subsection{The HI 21~cm line}
There is atomic hydrogen in our models,
suggesting that the clouds might be observed, in absorption or emission,
through the HI 21cm line. Indeed a number of detections have been made
(see eg Allen 2000 for references).

Lines at radio wavelengths are characterized
by their brightness temperatures, which will be nearly equal to the spin
temperature of the line at the position where it reaches an optical
depth of roughly unity. The clouds we model are quite optically thick
in the 21cm line ($\tau\sim10^3$ for the standard cold cloud computed
here) so the observed brightness temperature will depend on which side of the
cloud is viewed.  An observer viewing the cloud from its illuminated
face would see a brightness temperature of $\sim15$ K,
while an observer viewing the shielded side would see a slightly lower
temperature, nearer 10 K.  Note that these would be the observed
brightness temperatures if the clouds fully fill the telescope's beam.
If the clouds do not fill the beam the observed brightness temperature
will be lower as the remainder of the beam
sees the cosmic microwave background. Our model makes no prediction
for the cloud filling factor.

\section{Summary}
We have shown that a pressure-confined cloud at a radius of about
100~kpc in a cooling flow irradiated by a cooling flow has a large
cold core at a temperature of less than 17~K where cooling by atomic
carbon is dominant. This coolant explains the major difference between
our result and that obtained using molecular cloud cooling rates by
O'Dea et al (1994) and others. Clouds with a small amount of
microturbulence, or dust, are colder
still. The temperature profile of the dusty cloud
drops very rapidly towards that of the microwave background.

Due to the continuing revisions of the atomic and molecular cross sections,
rates and processes that have taken place since FFJ, there have been changes
in our computed model which we have explored in detail. Further small changes
will probably occur over the next few years. 

We do not argue here that such clouds necessarily are the sink of the
cooled matter in cooling flows, or are the source of the excess X-ray
absorption inferred in their spectra. That will be explored elsewhere.
Our purpose has been to confirm, from a detailed calculation of the
thermal and radiation balance of X-ray irradiated and
pressure-confined gas, that very cold clouds
can be expected in the cooling flow environement.

\section{Acknowledgements}
We thank Mark Voit for comments on this and an earlier version
of this work.
Research in Nebular Astrophysics at the University of Kentucky is
supported by the NSF (AST 00-71180) and NASA (NAG 5-4235). ACF thanks
for Royal Society for support.

\end{document}